\documentclass[cameraready]{Interspeech}

\title{The Unseen Threat: How Complementary Combinations of Acoustic Features Compromise Speech Privacy}

\author[affiliation={1,2}, orcid=0009-0000-6687-7559, correspondingauthor]{Jule}{Pohlhausen}
\author[affiliation={1}, orcid=0000-0002-0595-0262]{Joerg}{Bitzer}

\address{
    $^1$ Institute of Hearing Technology and Audiology, Jade University of Applied Sciences, Oldenburg, Germany\\
    $^2$ Dept. of Medical Physics and Acoustics, Carl von Ossietzky Universität Oldenburg, Germany
}

\email{jule.pohlhausen@jade-hs.de}

\keywords{speech privacy, privacy evaluation, acoustic feature fusion, speech recognition, privacy-utility trade-off}

\usepackage{csquotes}
\usepackage{subcaption}
\usepackage{stfloats}
\usepackage{multirow}
\usepackage{tabularx}

\begin{document}

\maketitle

\begin{abstract}

The interaction of acoustic features in leaking private information remains underexplored. This work focuses on time-domain features, i.e., broadband RMS level and pitch, alone and combined with temporally smoothed and subsampled Mel spectrograms.
Privacy is assessed using informed automatic speech recognition and speaker-verification attacks, while utility for conversational analysis is evaluated using voice activity detection and speaker diarisation.
RMS level and pitch alone exposed limited information, but fusion with \SI{500}{\milli\second}-smoothed Mel spectrograms enhanced utility while substantially increasing speech-content leakage. Specifically, the combination with RMS level showed that fine-grained temporal dynamics can amplify leakage from coarsely sampled spectral features. Speaker-identity protection remained limited for all tested features. 
Overall, these findings indicate that privacy assessments performed on individual acoustic features may underestimate leakage after feature fusion.

\end{abstract}

\section{Introduction}
Audio recordings are increasingly used to derive meaningful insights into scenarios, activities and environments that characterise everyday life \cite{mehl2017ear,cummins2023multilingual, kim_advances_2023, boovaraghavan2024kirigami}. 
Following the principles of ecological momentary assessment (EMA), long-term recordings from lightweight, portable devices enable the analysis of conversational behaviour \cite{milek_eavesdropping_2018, harari_sensing_2020} or language development \cite{romeo_beyond_2018}. However, these power-constrained devices impose limits on computational capacity, memory, and energy consumption. As a result, real-time, on-device analysis of long-term recordings remains computationally infeasible for such recording systems.

In particular, the recording of spoken language raises privacy concerns, since speech signals encode sensitive personal information, such as the speaker identity, health status, political views, and gender \cite{nautsch2019speech_privacy,Krger2020}. Therefore, the European general data protection regulation (GDPR) and related legal frameworks require robust privacy safeguards to protect sensitive speech data from misuse. 
In this article, speech privacy is assessed along two dimensions -- linguistic content and speaker identity -- by employing an automatic speech recognition (ASR) system and an automatic speaker verification (ASV) system, respectively. Simultaneously, utility for conversational analysis is measured by applying a voice activity detection (VAD) model and a speaker diarisation (SD) system.

\begin{figure}[!b]
    \centering
    \includegraphics[width=0.96\columnwidth, trim={0.2cm 0.08cm 0.1cm 0.1cm},clip]{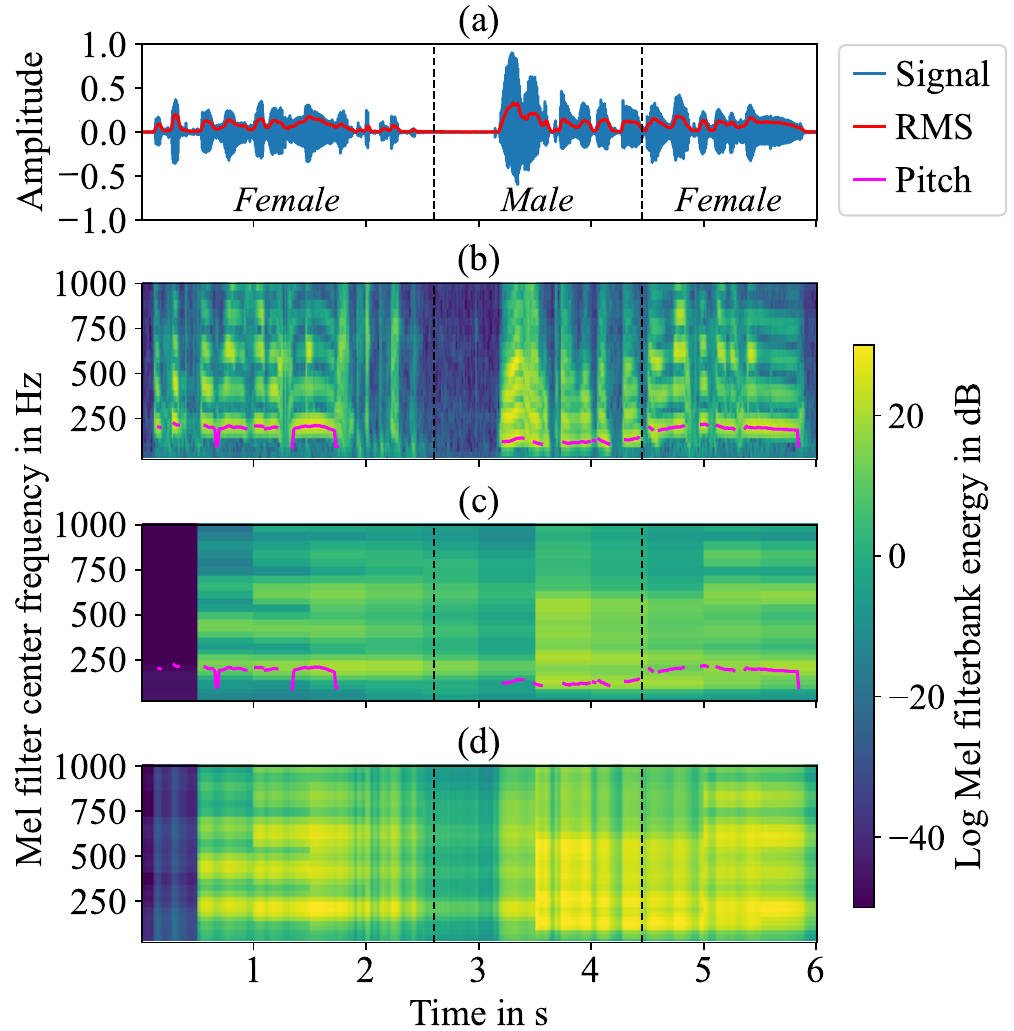}
    \caption{Acoustic features from a segment of meeting EN2002a  \cite{carletta_unleashing_2007}: (a) Time signal with RMS as red line, (b) original Mel spectrogram, (c) smoothed and subsampled ($\tau = 500$~ms) Mel spectrogram, and (d) same as in (c), modulated by RMS level. In (b) and (c), the magenta line shows the estimated YIN \cite{yin2002} pitch. Dashed vertical lines indicate speaker changes.} 
    \label{fig:feats}
\end{figure}

While raw audio is inherently sensitive, encryption or distributed learning are established approaches for privacy-preserving recording \cite{backstrom2025privacy}. 
This article shifts focus to the privacy risks associated with acoustic features in settings where feature extraction on power-constrained recording devices is separated from offline analysis with comparatively unconstrained computational resources.
In particular, Mel spectrograms are standard input features in speech processing pipelines due to their effective time-frequency representation and achieve state-of-the-art performance in diverse speech tasks \cite{speechbrain}. In order to enforce privacy protection of Mel spectrograms, temporal smoothing with subsampling \cite{Pohlhausen2025csl} is one possible technique with low computational complexity.
Specifically, a smoothing time $\tau = 500$~ms showed strong protection of spoken content (word error rate [WER] $> 100$~\%) but weak protection of the speaker identity (equal error rate [EER] of approx. 9~\%) when using an informed, retrained ASR and ASV model, respectively. Simultaneously, the utility performance, measured by VAD and SD, was slightly degraded \cite{Pohlhausen2025csl}. For a short speech signal, Figure~\ref{fig:feats} compares the original, unprotected Mel spectrogram in (b) 
\noindent
with the smoothed and subsampled version (c). This example illustrates that applying temporal smoothing with $\tau = 500$~ms blurs not only single words but also speech pauses and speaker changes.

Based on these findings, this study aimed to determine whether combining low-level acoustic features with temporally smoothed Mel spectrograms improves the analysis of conversations, and to what extent such combinations introduce privacy leakage.
First, it was anticipated that the block-wise computed broadband root mean square (RMS) level would increase the time resolution and thereby improve the accuracy of VAD decisions. As demonstrated in Figure~\ref{fig:feats}a, this high-resolution RMS preserves the overall dynamics of the temporal envelope. Modulating the \SI{500}{\milli\second}-smoothed Mel spectrogram with the corresponding RMS level, as shown in Figure~\ref{fig:feats}d, restores the envelope information.
However, the RMS is sensitive to background noise, which can diminish its informative value for VAD decisions \cite{graf_features_2015}.

Second, it was anticipated that pitch would enhance speaker change detection \cite{Hogg2021pitch} without revealing the speakers' identities. This hypothesis presupposes that, for a given group of speakers, each speaker is characterised by a different pitch distribution. For instance, the turns between a female and male speaker in Figure~\ref{fig:feats} correspond to changes in pitch.
Moreover, pitch and its variability are valuable for health-related speech assessments, such as monitoring the severity of depression \cite{Wadle2024depression}.

Due to the diversity of speech characteristics, a combination of complementary features is desirable to enhance the performance of both VAD \cite{van2013robust, graf_features_2015} and SD \cite{friedland2009prosodic, woubie2015using}.
Therefore, this article investigates whether RMS or pitch improve VAD or SD when combined with smoothed Mel spectrograms.
Simultaneously, it explores their potential to leak private information when utilised by informed \cite{Srivastava2020voice_conversion_attacks} attacker models (see Section~\ref{sec:pu_models} for details on attacker knowledge). 

The main contributions of this research are:
(1) A systematic evaluation of the linguistic content and speaker identity information revealed by RMS level and pitch, both individually and in combination with temporally smoothed and subsampled Mel spectrograms;
(2) A comprehensive assessment of the privacy-utility trade-off concerning conversational behaviour in everyday life, highlighting that complementary combinations compromise speech privacy.

\renewcommand{\arraystretch}{1.1}
\begin{table*}[tb]
\caption{Training and testing specifications using SpeechBrain \cite{speechbrain}.}
\label{tab:params}
\centering
\begin{tabular}{l|>{\raggedright\arraybackslash}p{1.8cm}>{\raggedright\arraybackslash}p{1.1cm}>{\centering\arraybackslash}p{3cm}>{\centering\arraybackslash}p{0.8cm}>{\centering\arraybackslash}c|>{\centering\arraybackslash}p{3.6cm}}
\toprule
Task $\downarrow$ & Architecture & Nr. of param. & Training data & Batch size & Epochs & Testing data    \\
\hline
\textbf{ASR}  & Transformer + CTC \cite{CTC} & 71.5M     & LibriSpeech 960~h \cite{librispeech} & 32 & 90 & LibriSpeech test-clean \cite{librispeech} \\
\hline
\textbf{ASV}  & ECAPA- & \multirow{2}{*}{21.9M} & \multirow{2}{*}{LibriSpeech 360~h \cite{librispeech}} & \multirow{2}{*}{32} & \multirow{2}{*}{30} & VPC \cite{voicepriv2020} test \\
\cline{1-1} \cline{7-7}
\textbf{SD}   & TDNN \cite{ECAPA} & & & & & AMI eval \cite{carletta_unleashing_2007} \\
\hline
\textbf{VAD}  & CRDNN \cite{speechbrain} & 109K & LibriParty train \cite{speechbrain} &  2  & 100 & LibriParty eval \cite{speechbrain} \\
\bottomrule
\end{tabular}
\end{table*}

\section{Experimental setup}\label{sec:setup}
The privacy-utility trade-off is evaluated using the experimental framework\footnote{Code is available at https://github.com/ol-MEGA/ppca} 
described in \cite{Pohlhausen2025csl}. The specifications for training and testing the models are summarised in Table~\ref{tab:params}.
For full details on the model architectures and evaluation metrics, the interested reader is referred to \cite{Pohlhausen2025csl}.
All evaluation metrics are reported in Section~\ref{sec:pu_results} with confidence intervals at a confidence level $\alpha = 5$~\% estimated using speaker-level bootstrapping.

\subsection{Attack and utility models}\label{sec:pu_models}
Linguistic-content privacy is evaluated using an ASR system, where a high word error rate (WER) is assumed to indicate sufficient protection. The WER exceeds 100~\% when the ASR system generates insertions not found in the reference transcript, commonly resulting from hallucinations or misaligned word boundaries.
The transformer-based\footnote{https://huggingface.co/speechbrain/asr-transformer-transformerlm-librispeech} ASR architecture \cite{speechbrain, CTC} was selected for its public availability and strong performance on standard benchmarks, achieving a WER of 2.34~\% on LibriSpeech test-clean (cf. Figure~\ref{fig:puto}c).

An ASV system \cite{ECAPA} is employed to assess the difficulty an adversary encounters in identifying known, enrolled speakers. Hence, a high equal error rate (EER) indicates better protection of the speaker's identity, where an EER of 50~\% reflects random decision behaviour.

The utility required for analysing conversations is evaluated by using a VAD \cite{speechbrain} and a SD system \cite{dawalatabad_ecapa-tdnn_2021}. Specifically, Matthews correlation coefficient (MCC) equals 1 for perfect classification of speech vs. non-speech frames, whereas a low diarisation error rate (DER) corresponds to a successful assignment of relative speaker labels.

This contribution considers only \textit{informed} attackers, which refers to the strongest attack paradigm \cite{voicepriv2020, Panariello2024VPC}. This means that attackers have complete knowledge of the privacy-preserving feature extraction and can train or finetune models on processed data using the same feature extraction. This overcomes the mismatch between a ready-to-use pretrained model and the maliciously captured features, potentially enabling attackers to exploit private information.
Notably, WER and EER are used as empirical proxies for privacy leakage under the evaluated attack models and conditions; they do not constitute formal guarantees of privacy.

\subsection{Training and test data}\label{sec:pu_data}
The publicly available datasets used for training, development, and evaluation of the privacy-protecting features presented in Section~\ref{sec:pu_feats} are sampled at \SI{16}{\kilo\hertz} (cf. Table~\ref{tab:params}). 
Specifically, the extraction of ECAPA-TDNN embeddings \cite{ECAPA} was trained on the \SI{360}{\hour} training split of LibriSpeech \cite{librispeech}, which comprises 921 speakers. In contrast, the ECAPA-TDNN model in \cite{Pohlhausen2025csl} was trained on VoxCeleb~2 \cite{VoxCeleb2}, containing 5994 speakers. This reduction in dataset size allows for significantly lower computational resources and faster training, while maintaining the same trends in results (cf. Section~\ref{sec:pu_results}).
Based on these speaker embeddings, the speaker verification was evaluated on the same test subsets of LibriSpeech and VCTK \cite{yamagishi2019vctk} following the configurations of the VoicePrivacy Challenge (VPC) 2020 \cite{voicepriv2020}. The weighted average EER, denoted as $Avg.^W$ EER, is reported.
Further, SD performance was evaluated on the augmented multi-party interaction (AMI) meeting corpus \cite{carletta_unleashing_2007} with the standard \textit{Full-corpus-ASR} partition using the \textit{HeadsetMix} recording streams. The SD relies on oracle VAD information, which is extracted from manual ground truth annotations.

\subsection{Privacy-protecting features}\label{sec:pu_feats}%
Extending the privacy-protecting features proposed in \cite{Pohlhausen2025csl}, we considered (1) temporally smoothed and subsampled Mel spectrograms with a smoothing time of $\tau = 500$~ms \cite{bitzer2016privacy, Pohlhausen2025csl},
(2) broadband RMS level, (3) pitch, and (4) combinations of RMS level or pitch with the smoothed Mel spectrograms.
The unprotected baseline performance is defined using a Mel filterbank with 80 filters (denoted as 80~Mel) as the input feature. These Mel spectrograms are computed on time segments with a window size of \SI{25}{\milli\second} and a hop size $L$ of \SI{12.5}{\milli\second}. As a result, the baseline input features are characterised by high frequency and time resolution, which is required for many downstream tasks, such as ASR and ASV. 
To obtain smoothed Mel spectrograms, the power spectral density (PSD) was recursively smoothed and subsampled by a factor of $\tau / L= 40$ \cite{bitzer2016privacy, Pohlhausen2025csl}. Each subsampled frame was then repeated 40 times to match the original frame rate before Mel filtering and enable frame-wise combinations.

The broadband RMS level was calculated on the same time segments.
Additionally, pitch was estimated using the YIN\footnote{https://github.com/brentspell/torch-yin} algorithm \cite{yin2002}, where the pitch search range was limited between 80 and \SI{400}{\hertz}. The lower pitch boundary of \SI{80}{\hertz} was chosen to ensure a hop size of \SI{12.5}{\milli\second}. 
We acknowledge that very few speakers exhibit a pitch below \SI{80}{\hertz}, which will result in a negligible number of pitch segments being missed.

\subsection{Evaluation setup and hypotheses}\label{sec:pu_setup}%
\begin{figure}[!tb]
    \centering
   \includegraphics[width=0.85\columnwidth]{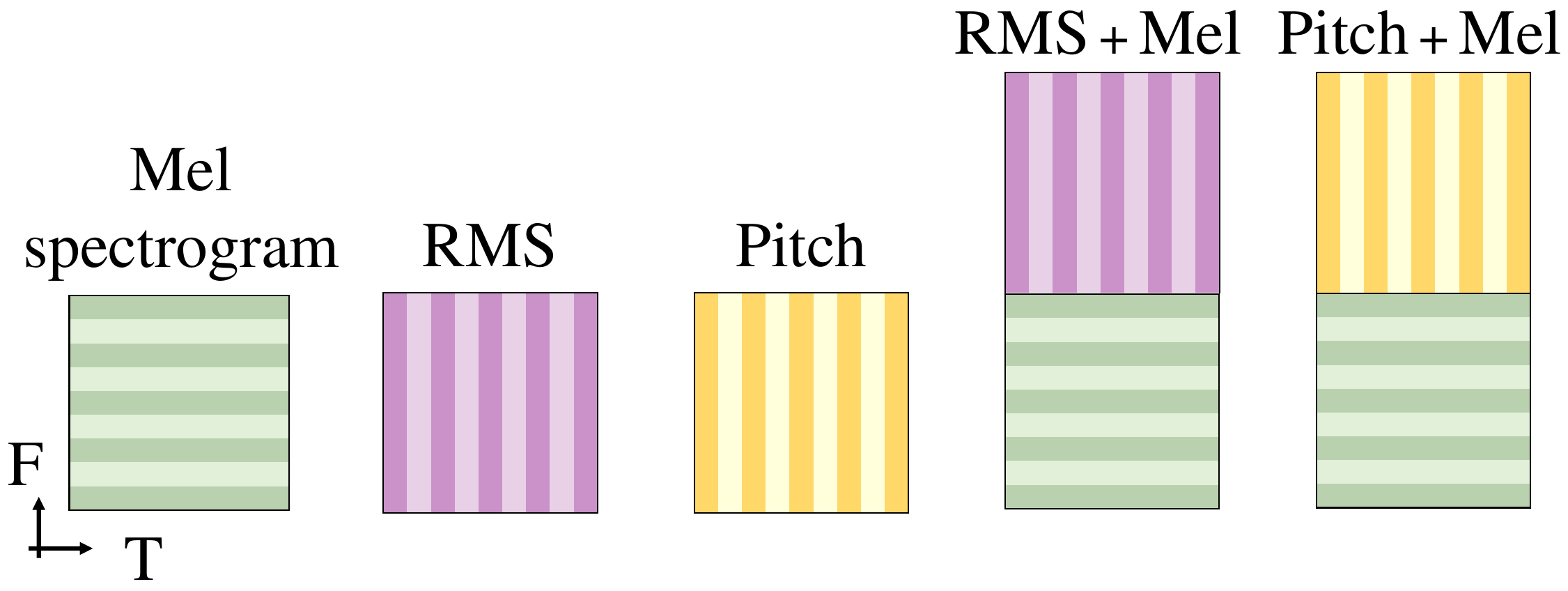}
    \caption{Dimensions of the considered features and their concatenations in the time-frequency (T-F) domain.} 
    \label{fig:feat_eval}
\end{figure}

Privacy and utility were evaluated under three scenarios, with separate informed models trained for each condition: (1) using only the RMS level, (2) using only the pitch, and (3) using a combination of temporally smoothed Mel spectrograms and either the RMS level or the pitch.
Input normalisation was applied separately to each feature. 
For scenarios (1) and (2), the one-dimensional RMS level and pitch were repeated 80 times for each time segment to match the input dimension of the baseline Mel filterbank, consisting of 80 filters. While this repetition adds redundant information, it ensures compatibility with the model architectures defined in \cite{Pohlhausen2025csl}. 
Both scenarios are expected to demonstrate the amount of private information contained in the RMS level and pitch solely. Figure~\ref{fig:feat_eval} provides an overview of the evaluated features and their combinations, which are detailed in the remainder of this section along with their hypothesised impact on privacy and utility performance.

We hypothesise that the RMS level will be suitable for correctly recognising short words but will likely fail to transcribe longer sentences. 
Analogously to \cite{pohlhausen_2024_iwaenc}, the robustness of the ASR performance using only the RMS level was assessed by adding noise from the Freesound portion of the MUSAN corpus \cite{musan, ko2017augment} to the test signals at signal-to-noise ratios (SNR) of \SI{10}{\decibel} and \SI{5}{\decibel}. It is expected that the ASR performance based on the RMS level will strongly degrade with decreasing SNR. 

Additionally, we expect that relying solely on pitch information will yield a high WER and precise but less sensitive 
VAD decisions. This is because pitch is primarily extracted from voiced speech segments, thereby resulting in unvoiced speech segments being missed. 
For instance, using the estimated YIN pitch directly as VAD, i.e., a frame with pitch is assumed to contain speech, results in a precision of 93.20~\% and a recall of 13.20~\% on the test split of LibriParty \cite{speechbrain}. Although this low recall contradicts accurate timing information, such as estimating the distribution of speech pauses or speaker turns, pitch can still be a suitable indicator for conversation scenarios in long-term recordings. 

Regarding ASV and SD, we hypothesise that speaker embeddings derived solely from the RMS level will be less speaker-specific, as they possibly capture only speech-pause ratios or rudimentary estimates of the speaking rate, thereby leading to high EER and DER. In contrast, relying solely on pitch may result in speaker confusion when comparing speakers with a similar pitch distribution. Nevertheless, for speakers with distinct voices, SD may still be possible to some extent.

In scenario (3), the temporally smoothed Mel spectrograms were concatenated with either the repeated RMS level or pitch for each time segment. In this context, the repetition could also be interpreted as an optional weighting of feature importance, where a repetition lower than the baseline input dimension of 80 might result in a lower importance assigned to a feature during training. A preliminary analysis compared different VAD model architectures with parallel training branches that concatenated outputs from different layers. However, the results indicated that the VAD performance did not differ significantly across the various architectures. Consequently, the most straightforward yet effective technique of feature concatenation was implemented for all models.
Compared with using only temporally smoothed Mel spectrograms, these combinations are expected to primarily improve VAD and SD performance. Moreover, in accordance with the preceding hypotheses, it is expected that the combination with the RMS level will, to some extent, improve the ASR performance. To further investigate this, another condition was tested, where the RMS level is smoothed and subsampled with $\tau = 250$~ms before concatenation.

\section{Experimental results and discussion} \label{sec:pu_results}

\begin{figure*}[!t]
 \includegraphics[width=1.01\textwidth, trim={0.15cm 0.07cm 0.17cm 0.07},clip]{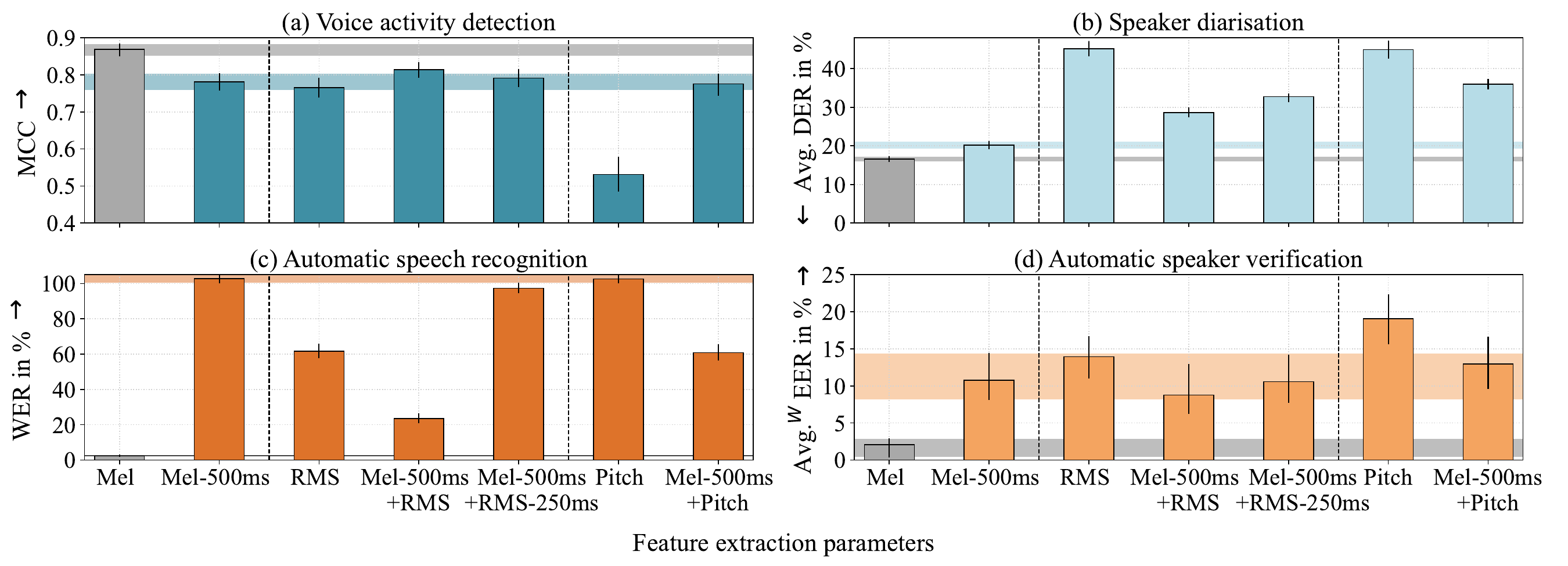}
 \caption{Privacy (ASR and ASV) and utility (VAD and SD) assessment for different input feature combinations. The first gray bar refers to the unprotected, original Mel spectrograms, respectively. 
 The 95~\% confidence intervals are represented by the vertical lines, respectively. Additionally, the shaded area indicates the confidence interval for the unprotected baseline performance (gray) and for the \SI{500}{\milli\second}-smoothed Mel spectrograms \cite{Pohlhausen2025csl}.}
\label{fig:puto}
\end{figure*}

For different input features and their combinations, Figure~\ref{fig:puto} summarises the evaluation metrics of the privacy and utility assessment. The arrows indicate the desired direction of performance change for each metric concerning the considered application scenario.
To enable direct comparison with previous assessments, the ASR and VAD results with the unprotected and the \SI{500}{\milli\second}-smoothed Mel spectrograms correspond to \cite{Pohlhausen2025csl} (see first two bars in Figure \ref{fig:puto}c and \ref{fig:puto}a). Notably, the differences in ASV and SD performance between training on \SI{360}{\hour} of LibriSpeech and VoxCeleb 2 \cite{Pohlhausen2025csl} are not statistically significant.

Employing solely the RMS level contains sufficient information to partially reveal spoken content, with a total WER of 61.67~\%. Specifically, 2.71~\% of all utterances in the test-clean subset of LibriSpeech \cite{librispeech} were completely correctly transcribed, including sentences with up to 22 words, thereby contradicting our initial expectations. Further, the training dataset contains only parts from these correctly transcribed utterances, but not the entire test utterance.
Over all speakers, the results indicate no clear trend that some speakers are generally easier to transcribe than others. 
Regarding the robustness against background noise, the total WER of the ASR model relying only on the RMS level degrades to 89.69~\% and 100.85~\% for an SNR of \SI{10}{\decibel} and \SI{5}{\decibel}, respectively. 
In contrast, the WER of an ASR model trained with the unprotected Mel spectrograms only slightly degrades to 5.74~\% and 10.81~\% for the same noise conditions \cite{pohlhausen_2024_iwaenc}.
This underscores the importance of the considered attack scenario, where the worst-case scenario for speech privacy involves recordings that are clean, with low reverberation.

As expected, the DER when relying solely on the RMS level is three times higher than the baseline SD performance, while the EER degrades only moderately compared to an EER of 50~\%, reflecting random decision behaviour. The latter observation was unexpected and demonstrates that for these clean recordings, even the RMS level can leak speaker identity. Moreover, this observation is not attributed to speaker-wise differences in the overall level, as all signals were normalised. We hypothesise that even the broadband RMS level partly encodes temporal prosodic cues, such as speech rhythm and speaking rate.

The VAD performance with the RMS level shows a similar degradation to that observed with \SI{500}{\milli\second}-smoothed Mel spectrograms. However, combining the high temporal resolution of the RMS level with the high frequency resolution of the Mel spectrogram improves the MCC.
For the same combination, the ASR performance decreases strongly to a total WER of 23.54~\%, highlighting the redundancy of speech. This unexpected strong improvement underscores the need for further privacy evaluation when multiple features are recorded, as their complementary combination can compromise speech privacy. 
Furthermore, this combination yields no significant ASV improvement and slightly degrades SD performance.

The combination of the \SI{250}{\milli\second}-smoothed RMS level with \SI{500}{\milli\second}-smoothed Mel spectrograms does not yield a statistically significant improvement in ASR and VAD performance. This implies that spoken content remains protected, but the expected gain in utility performance is not achieved.

Using solely pitch protects spoken content (total WER $>$ 100~\%), and yields a higher EER than using solely the RMS level. Both findings can likely be attributed to the omission of unvoiced segments. Consistent with the preliminary analysis in Section~\ref{sec:pu_setup}, the VAD model trained on pitch solely exhibits a precision of 92.49~\% and a recall of 67.70~\%, resulting in the lowest MCC of 0.53 compared to the remaining features.
Although pitch and RMS level yield similarly poor overall SD performance, the pitch-based model identifies two broad speaker clusters (primarily corresponding to female and male speakers) in meeting EN2002a, resulting in a DER of 26.09~\%. Consistent with prior work \cite{Hogg2021pitch}, this example illustrates that pitch-based diarisation strongly depends on the separability of the speakers' pitch distributions.

Finally, the combination of pitch with the \SI{500}{\milli\second}-smoothed Mel spectrograms improves primarily the ASR performance. For ASV and SD, this combination performs worse than using only \SI{500}{\milli\second}-smoothed Mel spectrograms, while VAD shows similar performance. Consequently, concatenating pitch information did not improve utility performance, but it remains an open question whether another fusion technique might lead to an improvement. Nevertheless, since the ASR performance compromises speech privacy to some extent, this aspect was not further investigated.

\section{Conclusions}\label{sec:con}
This article presents a systematic evaluation of speech privacy in time-domain acoustic features, focusing on the trade-off between privacy and utility for analysing the conversational behaviour in everyday life.
We employed a straightforward feature concatenation to combine broadband RMS level and pitch with temporally smoothed and subsampled Mel spectrograms.
As expected, the experimental results show that the RMS level alone partially exposes the linguistic content, and that pitch alone is insufficient for transcribing speech. Both low-level features offer marginally higher, albeit still limited, protection of speaker identity compared to smoothed Mel spectrograms. 
Regarding utility, only the combination of smoothed Mel spectrograms with RMS level enhances voicing decisions.
However, this utility improvement comes at a steep privacy cost: the fine-grained temporal dynamics of the RMS level complement the smoothed and subsampled Mel spectrograms, introducing a substantial privacy threat that is not captured when features are evaluated separately.
Under the evaluated clean-speech and informed-attacker conditions, these findings show that feature-wise assessments may underestimate privacy leakage after feature fusion. Privacy evaluations should therefore consider plausible combinations of recorded features. Further work should examine whether this leakage persists under naturalistic noise, reverberation, overlapping speech, alternative fusion strategies, and additional acoustic or multimodal feature combinations.

\ifcameraready
\section{Acknowledgements}
This work was supported by the Graduation program of Jade University of Applied Sciences (Jade2Pro 2.0).
\fi

\section{Generative AI Use Disclosure}
During the manuscript preparation, the authors used AI-assisted writing tools in order to improve language and readability. After using these tools, the authors reviewed and edited the content as required and take full responsibility for the content of the publication.

\bibliographystyle{IEEEtran}
\bibliography{sapstrings, mybib}

\end{document}